\documentclass[8.5pt,twoside,twocolumn]{article}

\usepackage[super,sort&compress,comma]{natbib} 
\usepackage{mhchem}
\usepackage{times,mathptmx}
\usepackage{sectsty}
\usepackage{balance} 
\usepackage{graphicx}
\usepackage{lastpage}
\usepackage[format=plain,justification=raggedright,singlelinecheck=false,
            font=small,labelfont=bf,labelsep=space]{caption} 
\usepackage{fancyhdr}
\begin{document}

\thispagestyle{plain}
\fancypagestyle{plain}{
\fancyhead[L]{\includegraphics[height=8pt]{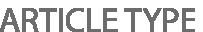}}
\fancyhead[C]{\hspace{-1cm}\includegraphics[height=20pt]{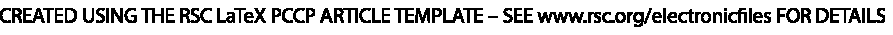}}
\fancyhead[R]{\includegraphics[height=10pt]{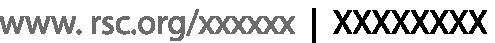}\vspace{-0.2cm}}
\renewcommand{\headrulewidth}{1pt}}
\renewcommand{\thefootnote}{\fnsymbol{footnote}}
\renewcommand\footnoterule{\vspace*{1pt}%
\hrule width 3.4in height 0.4pt \vspace*{5pt}} 
\setcounter{secnumdepth}{5}

\makeatletter 
\def\subsubsection{\@startsection{subsubsection}{3}{10pt}{-1.25ex plus -1ex minus -.1ex}{0ex plus 0ex}{\normalsize\bf}} 
\def\paragraph{\@startsection{paragraph}{4}{10pt}{-1.25ex plus -1ex minus -.1ex}{0ex plus 0ex}{\normalsize\textit}} 
\renewcommand\@biblabel[1]{#1}            
\renewcommand\@makefntext[1]%
{\noindent\makebox[0pt][r]{\@thefnmark\,}#1}
\makeatother 
\renewcommand{\figurename}{\small{Fig.}~}
\sectionfont{\large}
\subsectionfont{\normalsize} 

\fancyfoot{}
\fancyfoot[LO,RE]{\vspace{-7pt}\includegraphics[height=9pt]{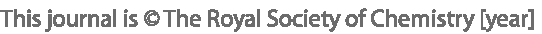}}
\fancyfoot[CO]{\vspace{-7.2pt}\hspace{12.2cm}\includegraphics{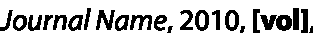}}
\fancyfoot[CE]{\vspace{-7.5pt}\hspace{-13.5cm}\includegraphics{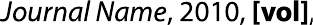}}
\fancyfoot[RO]{\footnotesize{\sffamily{1--\pageref{LastPage} ~\textbar  \hspace{2pt}\thepage}}}
\fancyfoot[LE]{\footnotesize{\sffamily{\thepage~\textbar\hspace{3.45cm} 1--\pageref{LastPage}}}}
\fancyhead{}
\renewcommand{\headrulewidth}{1pt} 
\renewcommand{\footrulewidth}{1pt}
\setlength{\arrayrulewidth}{1pt}
\setlength{\columnsep}{6.5mm}
\setlength\bibsep{1pt}

\twocolumn[
\begin{@twocolumnfalse}
\noindent\LARGE{\textbf{Structural band-gap tuning in g-C$_3$N$_4$}}
\vspace{0.6cm}

\noindent\large{\textbf{Sebastian Zuluaga,\textit{$^{a}$}
Li-Hong Liu,\textit{$^{b}$} Natis Shafiq,\textit{$^{b}$}
Sara Rupich,\textit{$^{b}$} Jean-Fran\c{c}ois Veyan,\textit{$^{b}$}
Yves J. Chabal,\textit{$^{b}$} and Timo Thonhauser$^{\ast}$\textit{$^{a}$}}}\vspace{0.5cm}

\noindent\textit{\small{\textbf{Received Xth XXXXXXXXXX 20XX, Accepted Xth XXXXXXXXX 20XX\newline
First published on the web Xth XXXXXXXXXX 200X}}}

\noindent \textbf{\small{DOI: 10.1039/b000000x}}
\vspace{0.6cm}

\noindent \normalsize{g-C$_3$N$_4$ is a promising material for hydrogen
production from water via photo-catalysis, if we can tune its band gap
to desirable levels.  Using a combined experimental and \emph{ab initio}
approach, we uncover an almost perfectly linear relationship between the
band gap and structural aspects of g-C$_3$N$_4$, which we show to
originate in a changing overlap of wave functions associated with the
lattice constants.  This changing overlap, in turn, causes the
unoccupied $p_z$ states to experience a significantly larger energy
shift than any other occupied state ($s$, $p_x$, or $p_y$), resulting in
this peculiar relationship.  Our results explain and demonstrate the
possibility to tune the band gap by structural means, and thus the
frequency at which g-C$_3$N$_4$ absorbs light.}
\vspace{0.5cm}
\end{@twocolumnfalse}]

\section{Introduction}
\label{intro}

\footnotetext{\textit{$^{a}$~Department of Physics, Wake Forest
University, Winston-Salem, North Carolina 27109, USA; E-mail:
thonhauser@wfu.edu}} \footnotetext{\textit{$^{b}$~Department of
Materials Science and Engineering, University of Texas at Dallas,
Dallas, Texas 75080, USA }}

Hydrogen production is a critical step in a possible future hydrogen
economy.\cite{Barreto_2003:hydrogen_economy,
Conte_2001:hydrogen_economy, Penner_2006:steps_toward,
Marban_2007:towards_hydrogen}  Water, due to its abundance, could be an
ideal source of hydrogen in the production process. Among the various
ways to split water, photocatalysis---using the sun as a source of
energy and a photocatalytic material---is one of the most promising
methods. \cite{Linsebigler_1995:photocatalysis_tio2,
Fujishima_2000:titanium_dioxide, Fox_1993:heterogeneous_photocatalysis,
Mills_1997:overview_semiconductor, Ni_2007:review_recent,
Li_2011:highly_efficient, Yan_2009:visible-light-driven_hydrogen,
Zou_2001:direct_splitting, Yeh_2010:graphite_oxide,
Zong_2008:enhancement_photocatalytic} However, the use of sunlight sets
two important restrictions on the photo-catalytic material used as
anode: First, it should be a semiconductor capable of absorbing light in
the visible range of the solar spectrum (between 1.6 and 3.2~eV).
Second, the H$^+$/H$_2$ and the O$_2$/H$_2$O electrode potentials should
lie in-between the edges of the conduction and valence bands. One of the
most studied semiconductors for this purpose is
TiO$_2$,\cite{Serpone_2006:is_band, Morikawa_2001:band-gap_narrowing,
Nagaveni_2004:synthesis_structure, Wang_2005:pyrogenic_ironiii-doped,
Zhu_2009:band_gap, Yin_2010:effective_band, Ong_2014:Self-assembly,
Qian_2014:Design, Xing_2014:Bi2Sn2O7-TiO2, Xu_2014:Photocatalytic,
Moon_2014:Decoration} however, its large band gap of 3.2~eV restricts
adsorption to the UV range of the
spectrum.\cite{Linsebigler_1995:photocatalysis_tio2} Thus, the search
for better suited materials is a highly active field of current
research.

Graphitic carbon nitride g-C$_3$N$_4$ is a layered system well known for
having a band gap of 2.7 eV.\cite{Thomas_2008:graphitic_carbon} It is
the most stable allotrope of carbon nitrides at ambient condition,
\cite{Xu_2012:band_gap} consisting of abundant elements.  It is an
efficient visible light photocatalytic
material,\cite{Dong_2011:efficient_synthesis, Yan_2009:Photodegradation,
Yan_2010:photodegradation_rhodamine} which exhibits promising properties
towards the reduction of CO$_2$ and other
pollutants.\cite{Min_2014:Enhanced, Yu_2014:Photocatalytic,
Katsumata_2013:Preparation, Maeda_2013:polymeric, Shi_2014:Polymeric} In
addition, g-C$_3$N$_4$ has shown remarkable properties toward the
photocatalytic splitting of water.\cite{Ran_Earth-abundant,
Martin_2013:H2-and-O2}  Also, the edges of the conduction and valence
bands are in the correct position with respect to the H$^{+}$/H$_2$
reduction and O$_2$/H$_2$O oxidation potentials.   X-ray diffraction
(XRD) shows that the distance between layers is $\sim$3.26~\AA, while
the in-plane lattice constant varies from 6.71 to
6.81~\AA.\cite{Wang_2009:metal-free_polymeric,
Liu_2011:simple_pyrolysis} Although the band gap is only slightly above
its optimal value, the photo-catalytic activity could significantly
benefit from a small band-gap reduction. To this end, g-C$_3$N$_4$ has
been doped with S,\cite{Liu_2010:unique_electronic,
Zhang_2011:sulfur-mediated_synthesis}
B,\cite{Yan_2010:photodegradation_rhodamine}
P,\cite{Zhang_2010:phosphorus-doped_carbon}
C,\cite{Dong_2012:carbon_self-doping} O,\cite{Li_2012:facile_approach}
Zn,\cite{Yue_2011:hydrogen_production} and
B--F.\cite{Wang_2010:boron-_fluorine-containing} Unfortunately, most
dopants do not show the desired effect; they increase the band gap,
leave it unaffected, or turn the system metallic.  Among those dopants,
the most promising is O. Li and co-workers
\cite{Li_2012:facile_approach} found that the O-doped g-C$_3$N$_4$ has a
band gap of 2.49~eV and consequently exhibits a better visible-light
photo-activity. Using X-ray photoelectron spectroscopy (XPS), Raman, and
Fourier transform infrared spectroscopy analysis, the authors were able
to determine that the introduced O atoms replace N atoms and their XRD
measurements showed that the peak located at 27.3$^{\circ}$ shifted to
27.5$^{\circ}$, indicating that the layer distance decreased from
3.26~\AA\ to 3.24~\AA.  While this is significant progress, O doping by
itself does not allow to tune the band gap to any desired level.
Furthermore, a clear understanding of what influences the band gap in a
predictive way is still lacking.  Motivated by the influence of O doping
on the layer separation, we pursue here a different route and explore
the relationship between the band gap and purely structural aspects of
g-C$_3$N$_4$.

In 2009, Wang et al.\cite{Wang_2009:metal-free_polymeric} found an
interesting---but as of yet unexplained---correlation between the
temperature at which their g-C$_3$N$_4$ samples were prepared and the
band gap: An increase in the temperature results in a decrease in the
band gap.  In the same work, the authors also report a change in the XRD
peak located at 27.3$^{\circ}$ towards higher angles as the temperature
increases. Similar results were found by Dong and
co-workers.\cite{Dong_2011:efficient_synthesis} This experimental
evidence strongly suggests a direct relationship between the layer
separation and the band gap in g-C$_3$N$_4$, but this relationship was
not explored in either of those works, as the focused was on sample
preparation and characterization. The investigation of this relationship
and its underlying physics is the main focus of the present article.

It is well known that several systems exhibit a relationship between
structural aspects and the band gap.\cite{Yang_1999:band-gap_change,
Ni_2008:uniaxial_strain, Minot_2003:tuning_carbon,
Sun_2008:strain_effect} Of particular interest is the similarly-layered
graphene monoxide, where the band gap can be tuned through
strain.\cite{Pu_2013:strain-induced_band-gap} In this case, the change
of the band gap was shown to be due to large variations in the
conduction band, while the valence band experiences only small changes
upon deformation of the system's geometry. This particular finding
guided our own research on g-C$_3$N$_4$.

\section{Experimental and Computational Details}
\label{comp_detail}
\subsection {Experimental Details}

In a typical synthesis, 2~g of melamine monomer (TCI) were put into an
alumina boat with a cover, and heated to a certain temperature in the
range of 500 -- 650~$^{\circ}$C with a ramp rate of
5~$^{\circ}$C$\cdot$min$^{-1}$ in a horizontal tube furnace (Carbolite
MTF3216) under flowing Ar, and then the temperature was kept at the
certain temperature (500, 550, 600, or 650~$^{\circ}$C) for 4 hours. The
resulting powders with color from light yellow to light brown---see the
inset in the lower panel of Figure \ref{XRD-BG_fig}---were collected for
use without further treatment. This change in color has been observed by
other groups and is believe to be due to the change in the layer
condensation and the change in the band gap of the
system.\cite{Martin_2013:H2-and-O2}

XRD diffraction patterns were recorded for  $2\theta$ values ranging
from 5$^{\circ}$ to 60$^{\circ}$ in order. X-ray diffraction
measurements were conducted with a Rigaku Ultima III diffractometer (Cu
K$\alpha$ radiation, X-ray wavelength of 1.54187~\AA, operating at
40~keV with a cathode current of 44~mA). For visualization purposes, a
running average over 20 data points has been made to plot the XRD data.
The UV-Vis spectra were obtained for the dry-pressed KBr disk samples
using Cary 5000 UV-Vis-NIR spectrophotometer. The spectra were recorded
from 200~--~800~nm at ambient conditions.

\subsection {Computational Details}

Our \emph{ab initio} calculations were performed at the DFT level with
the \textsc{Vasp} code.\cite{Kresse_1996:efficient_iterative} We used
projector augmented wave (PAW) pseudopotentials
\cite{Kresse_1999:ultrasoft_pseudopotentials} and a plane-wave expansion
with a kinetic-energy cutoff of 983~eV. Due to the strong van der Waals
interaction between the layers of g-C$_3$N$_4$, we used the vdW-DF
exchange-correlation functional.\cite{Dion_2004:van_waals,
Thonhauser_2007:van_waals, Langreth_2009:density_functional} In all
cases, we relaxed our systems until the forces on all atoms were less
than 1~meV/\AA.  We used a $k$-point sampling of $5\times 5 \times 5$
centered at the $\Gamma$ point.

To model g-C$_3$N$_4$, we constructed a hexagonal supercell with lattice
vectors of magnitude $a=b$ and $c$. The supercell contains two stacked
tri-$s$-triazine units in order to simulate possible stacking
configurations. Throughout this manuscript we will refer to the
separation between layers as $d$, which is exactly half the lattice
constant $c$, i.e.\ $d=c/2$.  Each tri-$s$-triazine unit is comprised of
eight N atoms and six C atoms---thus, our systems contain 28 atoms in
their supercell.

We used the $GW$ approximation at the level of $G_0W_0$ to calculate the
electronic structure of the system and its band
gap.\cite{Hedin_1965:new_method} The same approach has given excellent
agreement with experiments in our previous work on this
material.\cite{Stolbov_2013:sulfur_doping}  The Green's function and the
screened Coulomb interaction were calculated within DFT, where the
exchange energy was treated at the level of Hartree Fock. The Green's
function and the dielectric matrix were calculated using 140 bands,
while the cutoff energy for the response function was set to a value of
90~eV.

\section{Results}
\subsection{Layer Stacking and Band Gap}

\begin{figure}[t]
\includegraphics[width=\columnwidth]{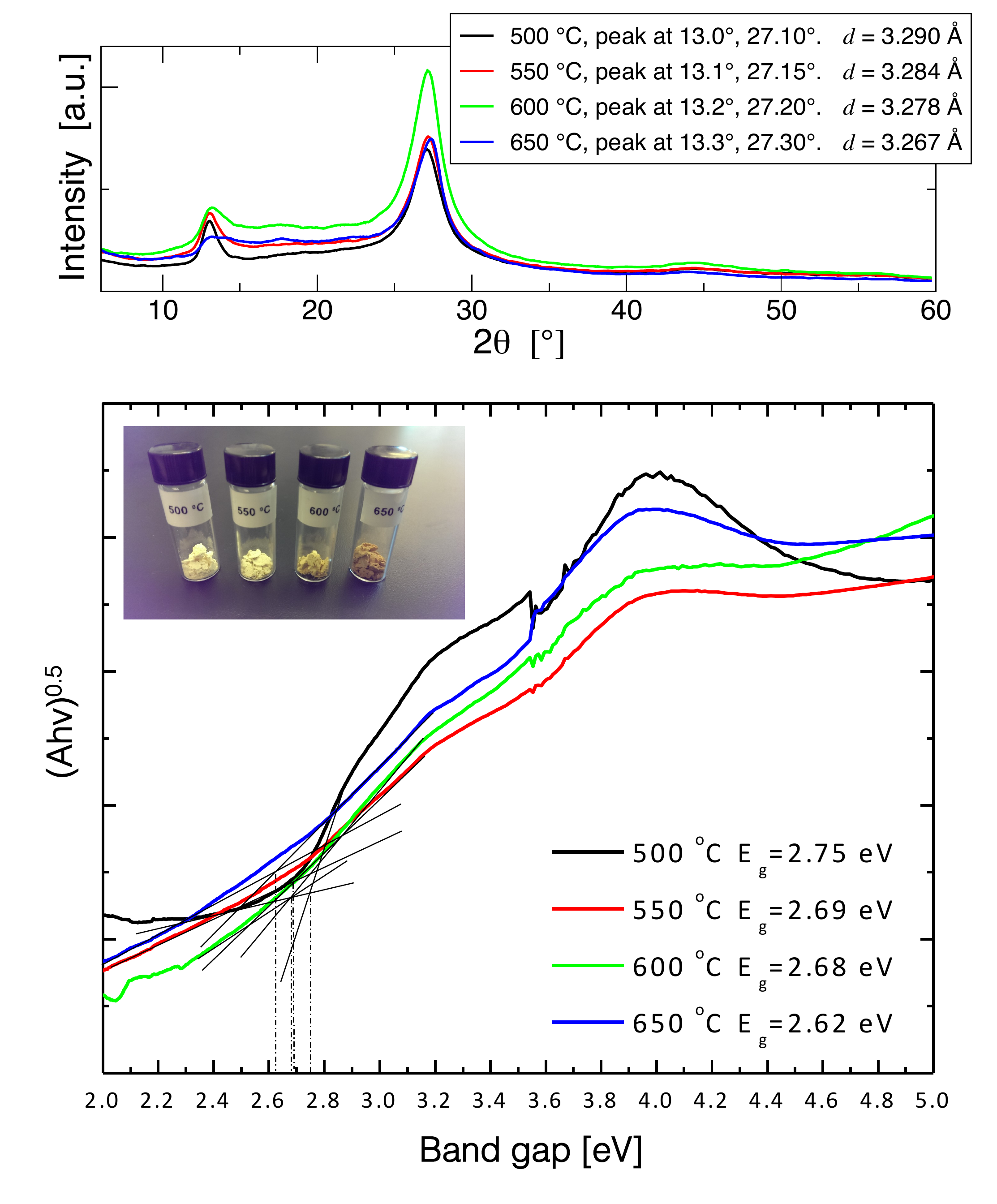} 
\caption{\label{XRD-BG_fig} (top) Experimental XRD pattern of the four
g-C$_3$N$_4$ samples prepared at 500, 550, 600, and 650~$^{\circ}$C.
(bottom) (Ahv)$^{1/2}$ vs.\ photon energy for the same four g-C$_3$N$_4$
samples. In the inserted figure the four g-C$_3$N$_4$ samples are shown.
The corresponding band gaps are shown in the legend.}
\end{figure}

We begin by reporting the experimental layer separation and band gap.
XRD measurements were taken on the four g-C$_3$N$_4$ samples prepared at
500, 550, 600, and 650~$^{\circ}$C. In addition, the band gap of these
samples was measured through the absorbance in the UV-Vis spectrum;
results are shown in Figure~\ref{XRD-BG_fig}.  The upper panel clearly
shows the peak at $\sim$27$^\circ$, which increases from 27.1$^\circ$ to
27.3$^\circ$ as the temperature increases from 500 to 650~$^\circ$C,
corresponding to a reduction of the inter-layer distance $d$ from 3.290
to 3.267~\AA. In the lower panel of this figure, the Tauc plot shows that
the band gap decreases from 2.75 to 2.62~eV over the same temperature
range. Using ultraviolet photoelectron spectroscopy (UPS) we were able to
determine that the valence bands of the samples prepared at 500, 550, and
600~$^\circ$C are located 0.9~eV below the Fermi level, while the valence
band of the sample prepared at 650~$^\circ$C is located 1.13~eV below
the Fermi level (see the supplementary materials for further details).
As mentioned earlier,
the behavior of the layer separation and band gap as a function of sample
preparation temperature has been observed separately
before,\cite{Li_2012:facile_approach,Wang_2009:metal-free_polymeric,
Dong_2011:efficient_synthesis} but a direct relationship between the
lattice constant and band gap was not considered.

To explain the above experimental results, further insight into the
structure of g-C$_3$N$_4$ is needed. Unfortunately, the XRD spectra do
not allow for a detailed structural analysis. While the large peak
around 27$^\circ$ is related to the layer separation and reveals
insightful information, the subtleties of the precise layer stacking is
encoded in a series of peaks at higher angles (see Figures~S1 and S2 in
the supplementary materials), which cannot be resolved experimentally.
Note that, in the upper panel of Figure~\ref{XRD-BG_fig} there is
another feature around 45$^\circ$, which we will analyze further below.
As such, we will use a combined experimental and theoretical approach to
shed light on the layer stacking as a function of temperature.

\begin{figure}[t]
\includegraphics[width=\columnwidth]{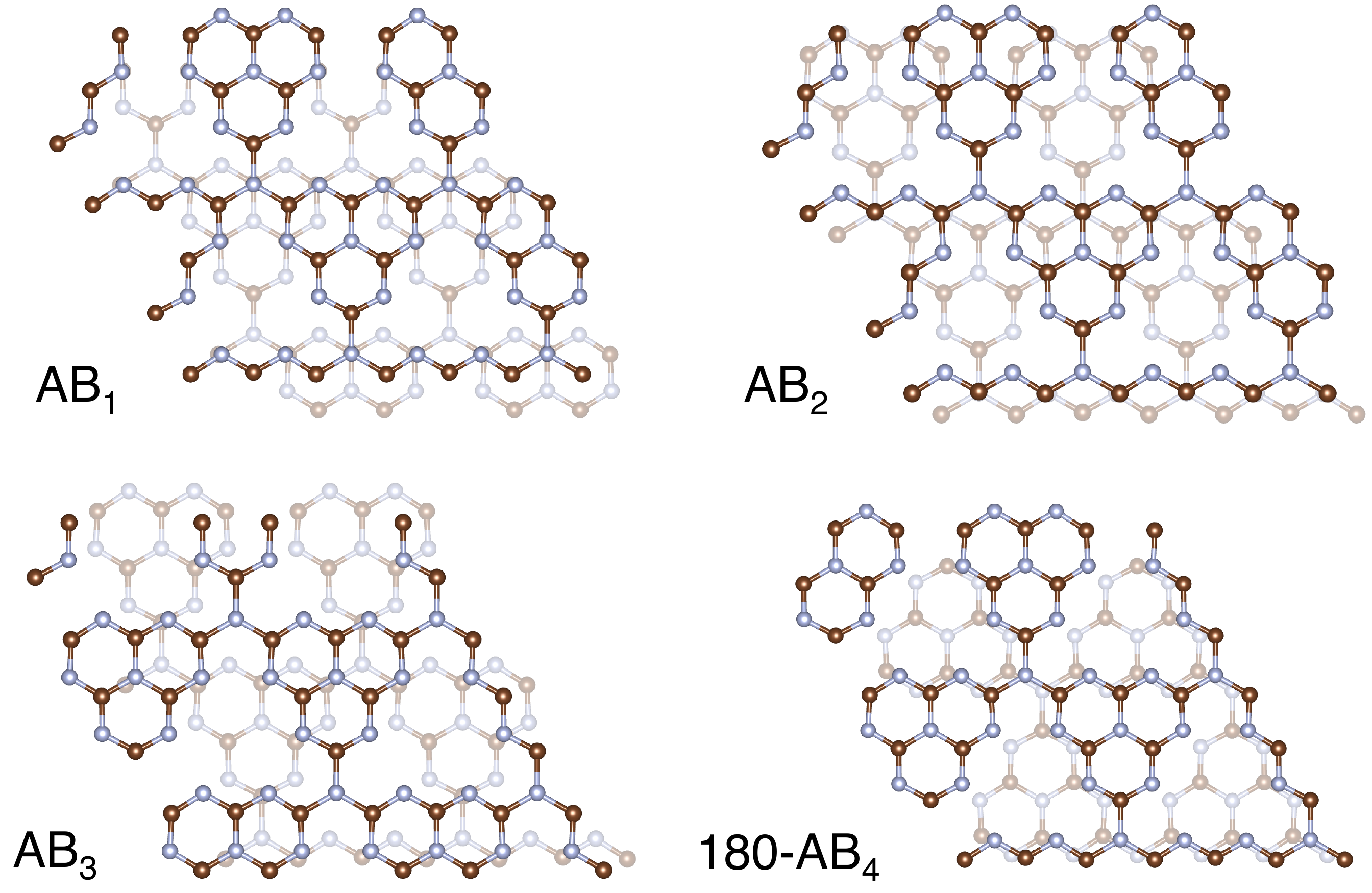} 
\caption{\label{stacking_geo_fig}
Layer stackings of g-C$_3$N$_4$ considered in this study.}
\end{figure}

The layer stacking of g-C$_3$N$_4$ exhibits several (meta)stable
configurations, located at local minima in the potential energy surface.
To find possible stackings, we performed a quasi-random structure
search, starting from more than
20 possible different stackings, which were subsequently relaxed.
Surprisingly, all starting configurations relax to one of only four
possible final stackings, which are depicted in
Figure~\ref{stacking_geo_fig}.  The family of AB stackings (AB$_1$,
AB$_2$, and AB$_3$) exhibits tri-$s$-triazine units in adjacent layers
that are shifted with respect to each other. The nomenclature 180-AB
refers to a situation where the tri-$s$-triazine unit in one layer is
rotated 180$^\circ$ with respect to the adjacent layer---in this family
we only find one member, which we label 180-AB$_4$. Interestingly,
stacking the tri-$s$-triazine units exactly on top of each other
(referred to as AA) is not amongst the local minima; it is very high in
energy (see Table~\ref{stacking-theo_tab}) and will not be considered
further here.

\begin{table}
\caption{\label{stacking-theo_tab} Calculated relative energy $\Delta
E$ in eV per unit cell between different stackings of g-C$_3$N$_4$, as
well as optimized lattice constants $a$, $b$, and $d=c/2$ in \AA\ and
band gaps $E_g$ in eV.  $a=b$ refers to the in-plane lattice constants,
while $d=c/2$ measures the distance between layers.} 
\begin{tabular*}{\columnwidth}{@{}l@{\extracolsep{\fill}}cccr@{}}\hline\hline
Stacking     & $\Delta E$ & $a=b$     &$d$     &$E_g$  \\\hline
AA           &0.86        &7.174      &3.632   &3.13   \\
AB$_3$       &0.14        &7.177      &3.313   &3.00   \\
AB$_1$       &0.12        &7.178      &3.301   &2.93   \\
AB$_2$       &0.12        &7.177      &3.296   &2.93   \\
180-AB${_4}$ &0.00        &7.178      &3.297   &2.86   \\\hline\hline
\end{tabular*}
\end{table}

\begin{figure}[t]
\includegraphics[width=\columnwidth]{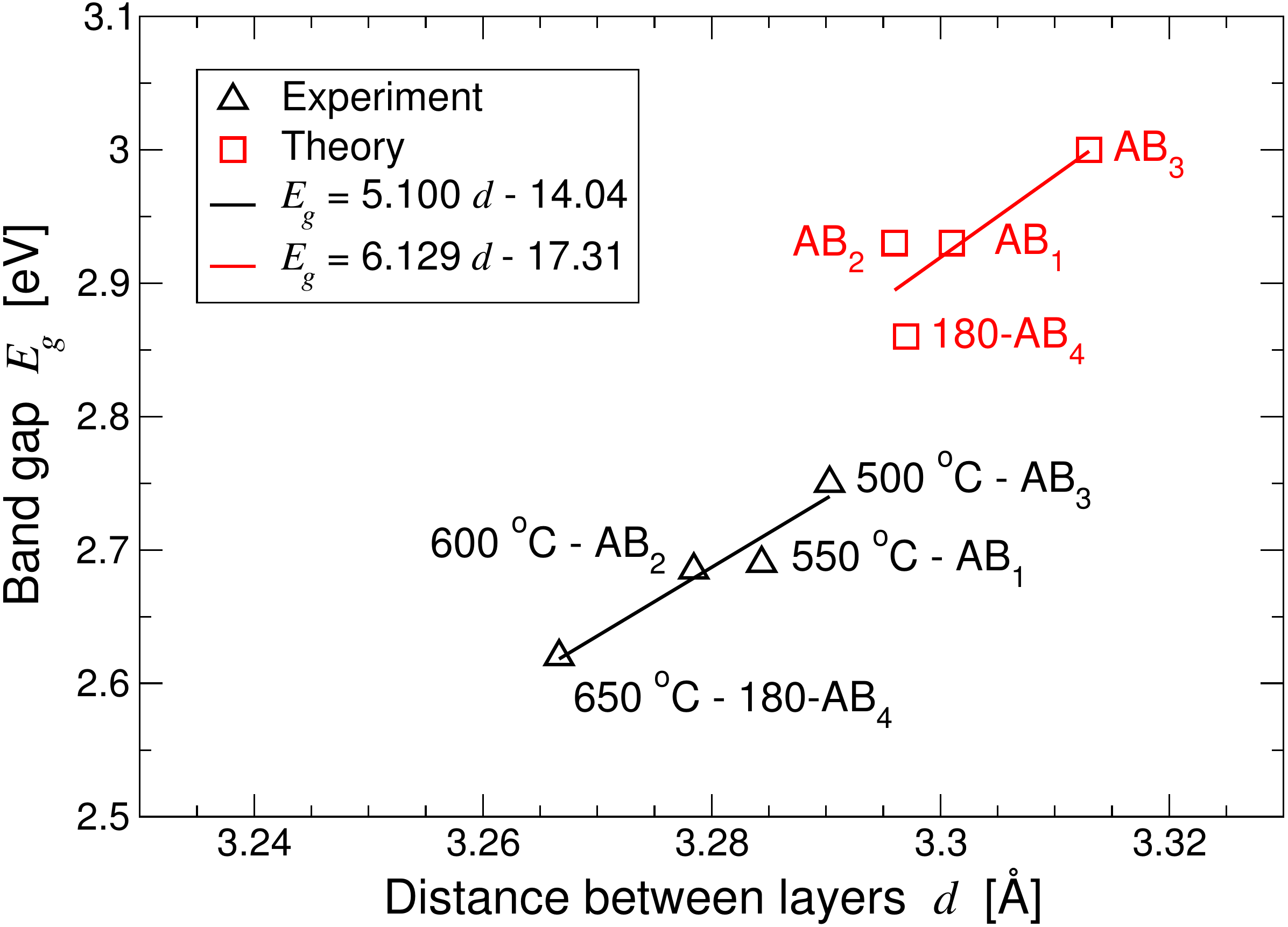} 
\caption{\label{exp-theory-BG-LC_fig} Band gap $E_g$ vs.\ separation of
layers $d$. Linear fits have been included for both the experimental and
theoretical results.} 
\end{figure}

Table \ref{stacking-theo_tab} shows the lattice constants, relative
energy, and band gaps of the different stacking configurations. As
expected, the lattice constants $a$ and $b$ are almost not affected by
the stacking configuration. Our values of $a$ and $b$ differ slightly
from our own XRD measurements and other
groups,\cite{Wang_2009:metal-free_polymeric} which indicate
$a=6.81$~\AA, however, yet other groups also report
$a=7.3$~\AA.\cite{Bojdys_2008:ionothermal_synthesis} On the other hand,
our results are in good agreement with the dimensions of the
tri-$s$-triazine unit, i.e.\
$a\simeq$7.13~\AA.\cite{Wang_2009:metal-free_polymeric} Overall, we find
the stacking 180-AB$_4$ to be lowest in energy, band gap, and layer
separation, and a direct relationship between those quantities becomes
apparent.

We combine our experimental and theoretical results in
Figure~\ref{exp-theory-BG-LC_fig}, which is one of the pertinent results
of our study. The two groups of points show a remarkable resemblance
(with the exception of the layer separation of the 180-AB$_4$ structure,
being predicted slightly too large), based on which we are the first to
assign stacking labels---and thus detailed structural information---to
the experimental results. This result suggests that the temperature at
which the sample is prepared determines the stacking configuration that
dominates the sample.  It is important to mention that, although we find
excellent agreement between experiment and theory concerning the change
in the band gap as a function of layer distance, there is a discrepancy
in the absolute values of the band gaps of $\sim$0.2~eV.  This
discrepancy is partly the result of vdW-DF's well-know tendency to
overestimate lattice
constants.\cite{Thonhauser_2006:interaction_energies} If we calculate
the band gap for the experimental lattice constants of e.g.\ 180-AB$_4$
($a=b=6.81$ and $d=3.267$~\AA), we find a band gap of 2.73~eV, which is
within 4\%\ of the experimental value. Note that with the $GW$ method we
calculate the quasi-particle gap, while our experiments measure the
optical gap, and their difference is related to the exciton binding
energy. As such, the remainder of the discrepancy is likely due to
excitonic effects, which Wei and co-workers have found to be important
in related systems.\cite{Wei_2013:Strong}

The assignment of the dominating stacking configuration to the
experimental data in Figure~\ref{exp-theory-BG-LC_fig} is also supported
by Figures~S1 and S2 of the supplementary materials, where we analyze the
peak in the XRD spectra around 45$^\circ$ further.  Figure~S2 of the
supplementary materials shows that all of our calculated stacking
configurations exhibit two peaks at 43.7$^{\circ}$ and
45.8$^{\circ}$ and that their heights gradually shift from the latter to
the former with decreasing total energy. As the two peaks cannot be
resolved separately in experiments, this manifests itself in a slight
overall shift  to lower angles of the broad peak measured, which is
exactly what is observed experimentally in Figure~S1 in the supplementary
materials.

Although not the main focus of this paper, one interesting question
remains---why do the samples assume different stacking configurations
when prepared at different temperatures?  We see from
Table~\ref{stacking-theo_tab} and Figure~\ref{exp-theory-BG-LC_fig} that
increasing temperatures leads to lower-energy stackings. It is tempting
to assume that the local energy minima of the higher-energy stacking
configurations are ``larger'' (and thus easier to get trapped into) and
that lower-energy stackings are separated by increasingly larger
barriers that can only be overcome by sufficiently high temperatures.
However, from our random-structure search we find that the first
statement is not true. And, from nudged-elastic band calculations to
find the transition states between stackings, we further find that the
second statement is not true either---the energy barriers separating
stackings are of such magnitude (on the order of 1 $k_BT$) that they can
easily be overcome at room temperature. As such, further research is
needed to answer this question and at the moment we are performing
\emph{ab initio} molecular-dynamics simulations to study the behavior of
this system as a function of temperature.

\subsection{A Model System}
\label{gap_lattice}

In the previous section we made the relationship between the layer
separation and the band gap explicit, providing even a linear equation
in Figure~\ref{exp-theory-BG-LC_fig}. However, we have not yet explained
the physical origin of this peculiar relationship, which is the main
focus of the present section. To this end, we use the stacking
configuration AB$_1$ as a model system and analyze this relationship
further. Note that we could have used either of the other stackings,
with only minimal changes in the results.

\begin{figure}[t]
\includegraphics[width=\columnwidth]{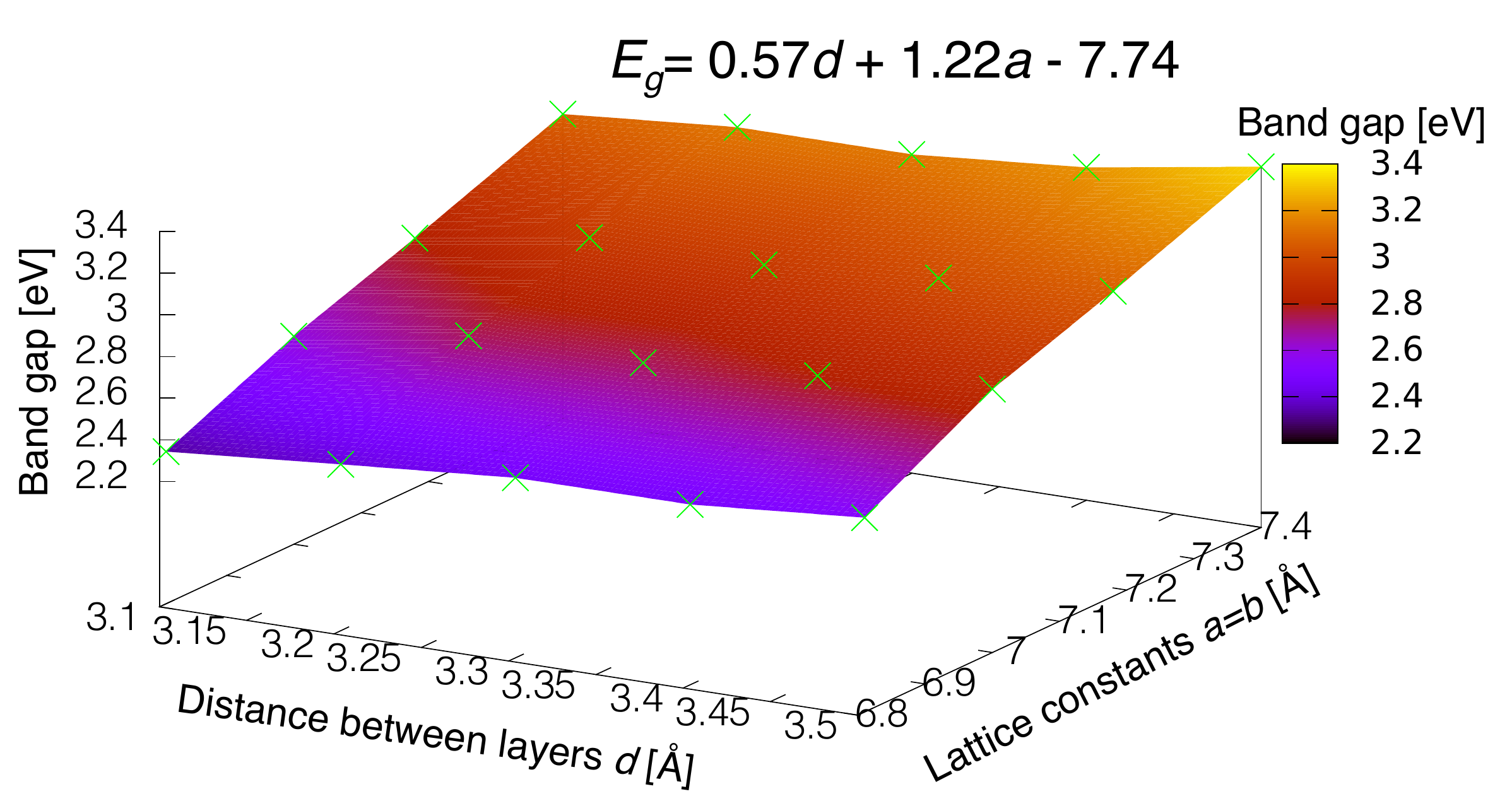}
\caption{\label{Bg_vs_Lc_fig} Band gap vs.\ lattice constants for the
g-C$_3$N$_4$ in the AB$_1$ stacking configuration. The green marks show
the points for which we have calculated the band gap. In between this
grid, the band gap is linearly interpolated.} 
\end{figure}

\begin{figure}[t]
\includegraphics[width=\columnwidth]{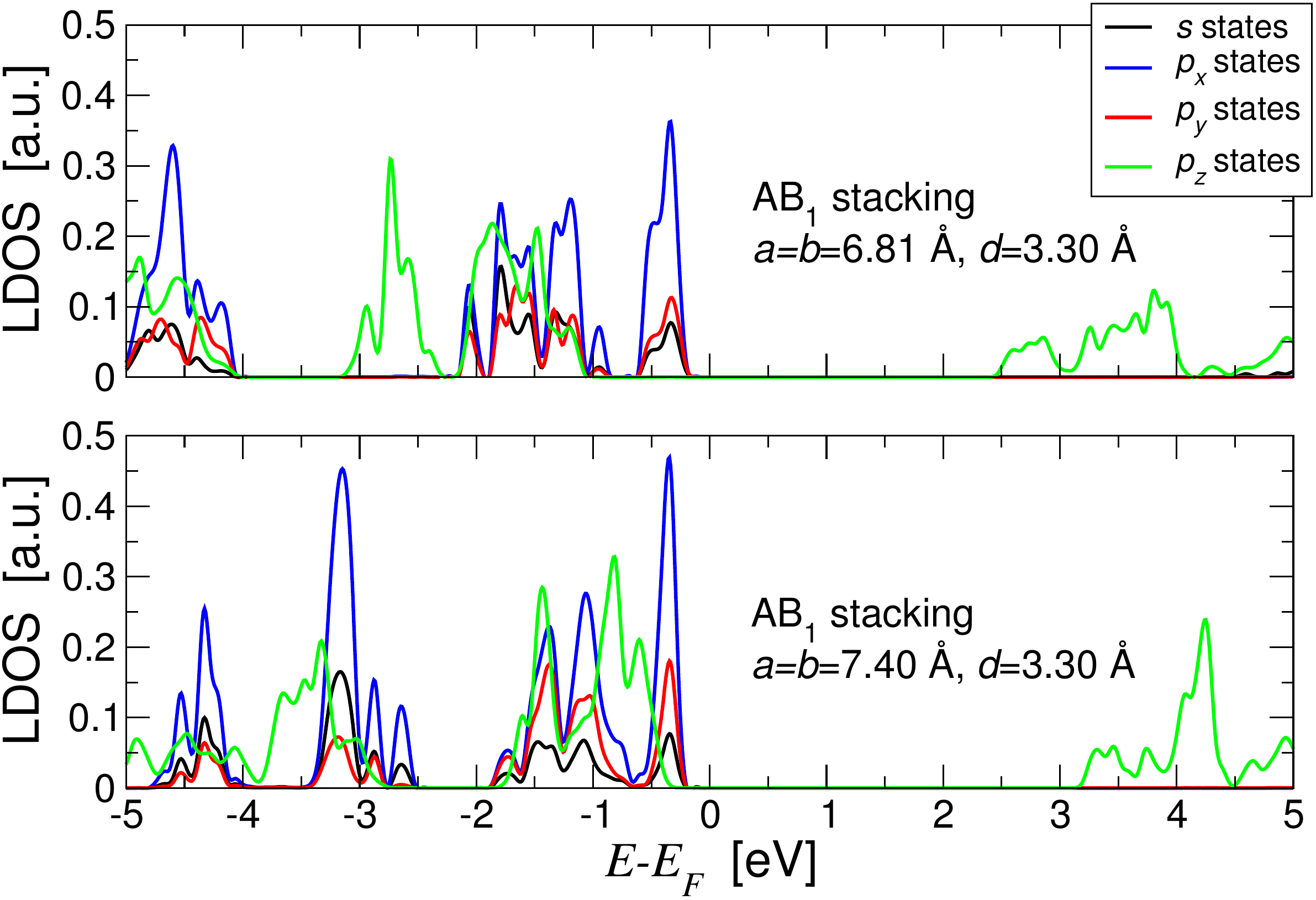}
\caption{\label{comparison_DOS_fig} LDOS of the N atom at the edge of
the tri-$s$-triazine unit, resolved into the states $s$, $p_x$, $p_y$,
and $p_z$. The upper and lower panel show the LDOS for different lattice
constants.} 
\end{figure}

First, we calculate the band gap of the system as a function of the
lattice constants $a=b$ and the distance between layers $d$; results are
presented in Figure~\ref{Bg_vs_Lc_fig}.  From this figure it is obvious
that the relation is almost perfectly linear in both dimensions and we
provide the equation of the corresponding fitted plane. To explain this
behavior, in Figure~\ref{comparison_DOS_fig} we plot the local density
of states (LDOS) of the N atom at the edge of the tri-$s$-triazine unit
for two different lattice constants. From the figure it can be seen
that, in both cases, the top of the valence band is comprised of $s$,
$p_x$, and $p_y$ states, while the bottom of the conduction band is
comprised only of unoccupied $p_z$ states. In particular, we find that
the unoccupied $p_z$ states shift towards higher energies as the lattice
constant $a$ increases from 6.81 to
7.40~\AA. In general, and true for all grid points calculated in
Figure~\ref{Bg_vs_Lc_fig}, the $p_z$ states experience a higher energy
shift than the $s$, $p_x$, and $p_y$ states, causing the band gap to
increase as the lattice constant $a$ increases. The exact same behavior
is found for increasing the layer separation $d$.

This is a well-known effect also observed in other
systems,\cite{Stolbov_2013:factors_controlling} due to the reduction of
hybridization between $p_z$ states of neighboring atoms, and we identify
it here as the main mechanism underlying the structure/band gap
relationship in g-C$_3$N$_4$.  To further support this conclusion, we
again use the AB$_1$ model system and calculate its band gap as the two
layers are rigidly slid agains each other from 0 to 0.6~\AA\ along the
unit vector $b$.  This way, the hybridization between atoms of adjacent
layers is forced to gradually grow. Indeed, we find that the conduction
band of the system is gradually displaced towards lower energies and the
band gap continuously changes from 2.93 to 2.88~eV. In summary, the
hybridization and overlap of the $p_z$ states, which clearly depends on
structural aspects such as the lattice constants, is the determining
mechanism in the structure/band gap relationship.

We conclude this section by estimating pressures necessary to lower the
band gap of g-C$_3$N$_4$. As can be seen in Figure~\ref{Bg_vs_Lc_fig},
the largest effect would result from reducing the lattice constants $a$
and $b$---but this is technologically not feasible, in particular in a
powder. On the other hand, from the same figure we know also the change
in the band gap with layer separation $d$, and we also know the total
energy as a function of $d$. As such, we can calculate the pressure
required to reach a particular layer separation and can then relate this
to the band gap, as can be seen in Figure~S3 of the supplementary
materials. Overall, we find that fairly large pressures are required and
the band gap changes by 0.03~eV for every GPa of pressure along the
c-axis. On the other hand, it may be possible to change the in-plane
lattice constant of g-C$_3$N$_4$ by growing it on a lattice-mismatch
substrate such as graphite, boron nitride, or metals with hexagonal
structures. Of particular promise is Sc, which exhibits the necessary
hexagonal symmetry with a lattice constant of
3.31 \AA\ when cleaved along the [0001] plane, potentially resulting in
a band gap of 2.51~eV (according to our fit from
Figure~\ref{exp-theory-BG-LC_fig}).

\subsection{Generalization to g-C$_3$N$_4$}

\begin{figure}[t]
\includegraphics[width=\columnwidth]{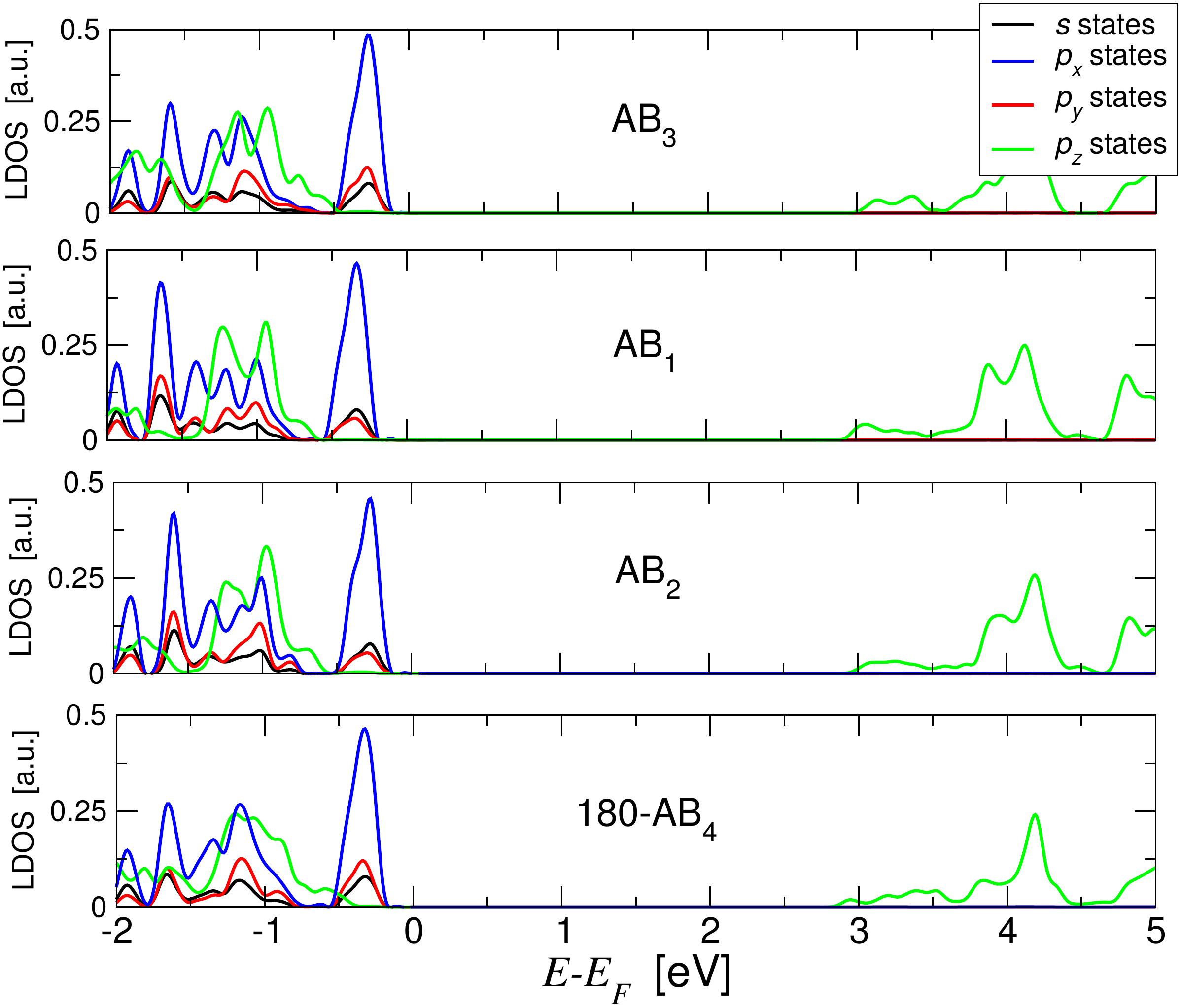}
\caption{\label{N8_stacking_comp_fig} LDOS of the N atom at the edge of
the tri-$s$-triazine unit. The LDOS has been resolved into the states
$s$, $p_x$, $p_y$ and $p_z$.}
\end{figure}

In the previous section we found that the hybridization of $p_z$ states
is responsible for the almost linear relationship between the lattice
constants and band gap in the model AB$_1$ system. However, these
results are easily generalized to the other stackings and in
Figure~\ref{N8_stacking_comp_fig} we plot the LDOS of the N atom at the
edge of the tri-$s$-triazine unit for the various stackings considered.
As expected, the conduction band in all four systems is comprised only
of $p_z$ states and the band gap is governed by the position of these
states.  In all cases the band gap is susceptible to stress and to the
stacking configuration adopted by the system through the same mechanism,
i.e.\ the conduction band of all stackings experiences large variations,
while the valence band experiences only small shifts upon changes in the
geometry of the system.

This mechanism also explains the findings of Li et
al.,\cite{Li_2012:facile_approach} where oxygen doping was found to
decrease the band gap of g-C$_3$N$_4$ by
0.21~eV. We now know that the influence of oxygen upon the electronic
structure and thus the band gap is mostly indirect, in that it changes
the layer separation from 3.26~\AA\ to 3.24~\AA\ and thus indirectly
influences the band gap through our proposed mechanism. This opens the
door for a different line of research, where the main focus is on
decreasing the layer separation by means of doping or intercalating
g-C$_3$N$_4$---a direction we are already actively pursuing.

\section{Summary}

In this work we combine experimental spectroscopy and \emph{ab initio}
calculations to study the g-C$_3$N$_4$ system. We have shown that the
band gap of g-C$_3$N$_4$ depends on the stacking configuration and, more
generally, it depends on the lattice constants in an almost perfectly
linear relationship. We explain this peculiar finding by analyzing the
density of states and find that the top of the valence band in this
system is comprised of $s$, $p_x$, and $p_y$ states, while the bottom of
the conduction band is comprised of $p_z$ states only. As the lattice
constants decrease, the increase of overlap between $p_z$ states of
neighboring atoms causes them to experience a higher energy shift than
the $s$, $p_x$, and $p_y$ states. As a consequence, the band gap
shrinks. 

Our main finding is the uncovering of this underlying mechanism that
controls the structure/band gap relationship. Knowledge and
understanding of this mechanism is essential for further band-gap tuning
of g-C$_3$N$_4$ for the purpose of photocatalysis, where it controls the
frequency of light absorbed and thus the efficiency of such devices. Our
main point is that band-gap tuning should focus on decreasing the
lattice constants of this system. Although we show that this is possible
through high external pressure, we believe that doping, intercalating,
or stress created by lattice mismatch is the most promising way to tune
the band gap of g-C$_3$N$_4$.

\section*{Acknowledgements}

This work was supported by Department of Energy Grant No.\
DE-FG02-08ER46491. NS, who performed the XRD measurements, was supported
by DOE grant No. DE-SC0010697. The authors would also like to thank
Dr.\ Weina Peng from the Department of Materials Science and Engineering,
University of Texas at Dallas, for her valuable discussions.

\footnotesize{
\bibliography{references,biblio}
\bibliographystyle{rsc}}

\end{document}



\title{Structural band-gap tuning in g-C$_3$N$_4$\\[1ex]
--- Supplementary Materials ---\\[0ex]\mbox{}}

\author{Sebastian Zuluaga}         \affiliation{\WFU}
\author{Li-Hong Liu}               \affiliation{\UTD}
\author{Natis Shafiq}              \affiliation{\UTD}
\author{Sara Rupich}               \affiliation{\UTD}
\author{Jean-Fran\c{c}ois Veyan}   \affiliation{\UTD}
\author{Yves J. Chabal}            \affiliation{\UTD}
\author{Timo Thonhauser}           \affiliation{\WFU}

\date{\today}

\maketitle

\onecolumngrid
\section{Ultraviolet photoelectron spectroscopy}
\twocolumngrid

Ultraviolet photoelectron spectroscopies (UPS) were obtained on a PHI
Versa~Probe~II scanning XPS microprobe, Physical Electronics, equipped
with a Prevac UV source (UV40A). The UPS radiation is generated by a
He-gas discharge lamp (He~I~=~21.22~eV; UV source settings:
$I_\text{emis} = 100$~mA, $U_\text{source}=0.52$~kV,
$P_\text{He}=9.6\times10^{-2}$~mbar), and measurements were made using
the He I excitation (21.2~eV) and recorded with a constant pass energy
of 1.75~eV in the ultrahigh vacuum chamber of the XPS instrument. Samples
were placed at normal incidence to the analyzer. The source to analyzer
angle was set to 45$^{\circ}$. During the measurement, both e-gun and Ar
gun were turned on for neutralization. The pressure in the main chamber
was $2.6\times10^{-7}$~Pa. Samples were loaded in the UHV chamber
overnight to allow sample degasing and to let the UHV pressure drop
below $1\times10^{-6}$~Pa.

\begin{figure*}[h]
\includegraphics[width=0.65\textwidth]{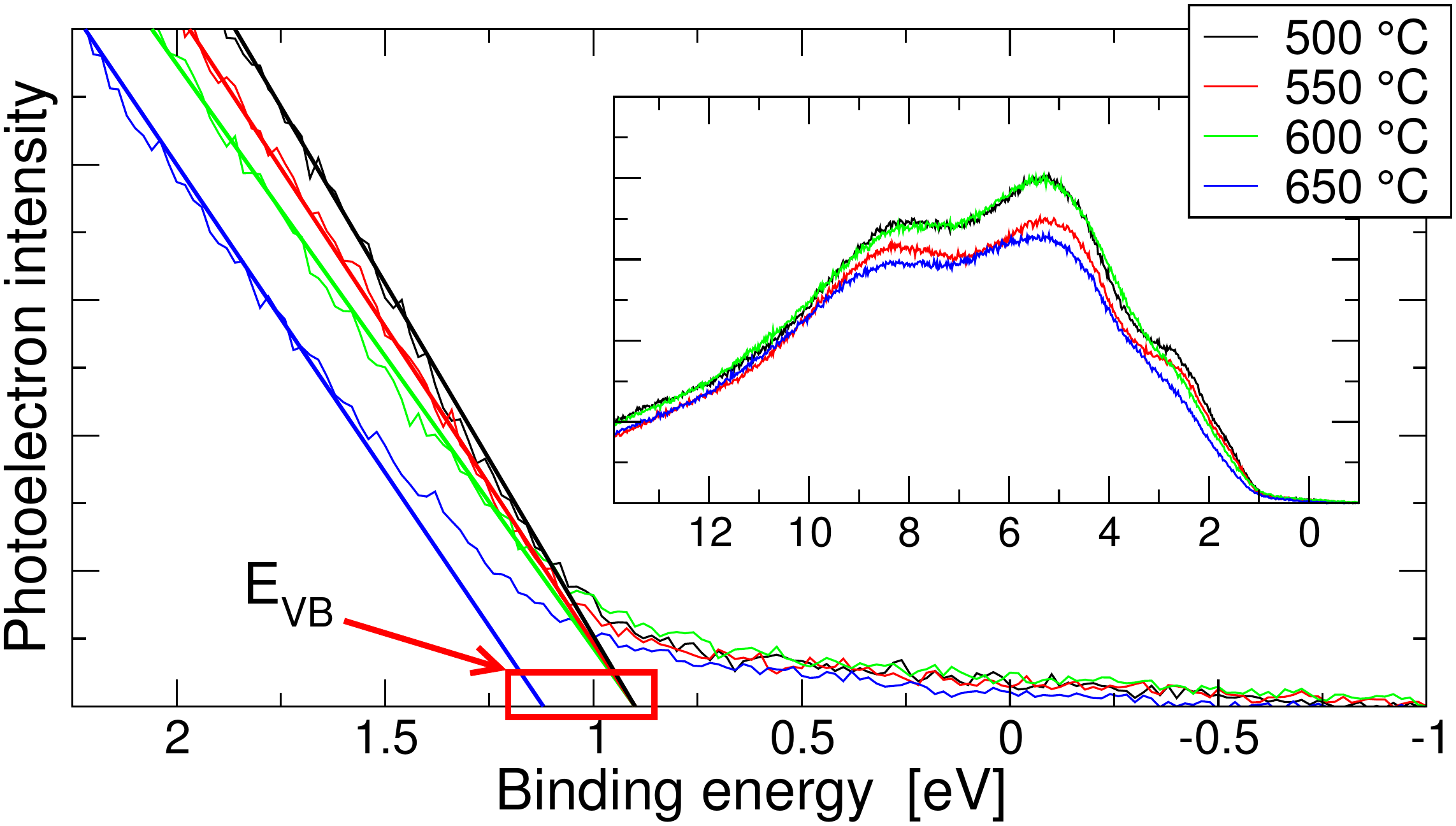}
\begin{minipage}{4.5in}
\caption{\label{UPS-sup_mat_fig}
UPS spectra in the valence band region of the four g-C$_3$N$_4$ 
samples prepared at 500, 550, 600, and 650~$^{\circ}$C. The inset 
shows the full spectra between $-1$ and 13~eV.}
\end{minipage}
\end{figure*}

\vspace{2.1cm}Figure \ref{UPS-sup_mat_fig} shows a series of UPS spectra
for the four g-C$_3$N$_4$ samples prepared at 500, 550, 600, and
650~$^{\circ}$C in the valence band region. In particular, the
$E_\text{VB}$ positions, based on linear extrapolation, are shown and
numerical values are collected in Table~\ref{UPS_tab}.  Because the
samples are insulators at room temperature, their work function cannot
be measured by standard bias techniques using UPS.

\begin{table}[h]
\caption{\label{UPS_tab} Band gap, $E_g$, and the positions of the
valence and conduction bands, $E_\text{VB}$ and $E_\text{CB}$, in eV
with respect to the Fermi level for the four g-C$_3$N$_4$ samples
prepared at different temperatures.} 
\begin{tabular*}{\columnwidth}{@{}l@{\extracolsep{\fill}}ccr@{}}\hline\hline
Temperature     &$E_g$ &$E_\text{VB}$     &$E_\text{CB}$      \\\hline
500 $^{\circ}$C       &2.75      &$-0.90$      &1.85 \\
550 $^{\circ}$C       &2.69      &$-0.90$      &1.79 \\
600 $^{\circ}$C       &2.68      &$-0.90$      &1.78 \\
650 $^{\circ}$C       &2.62      &$-1.13$      &1.49 \\\hline\hline
\end{tabular*}
\end{table}

\onecolumngrid
\newpage\section{Figures}

\vspace{0.25in}
\begin{figure*}[h]
\includegraphics[width=0.65\textwidth]{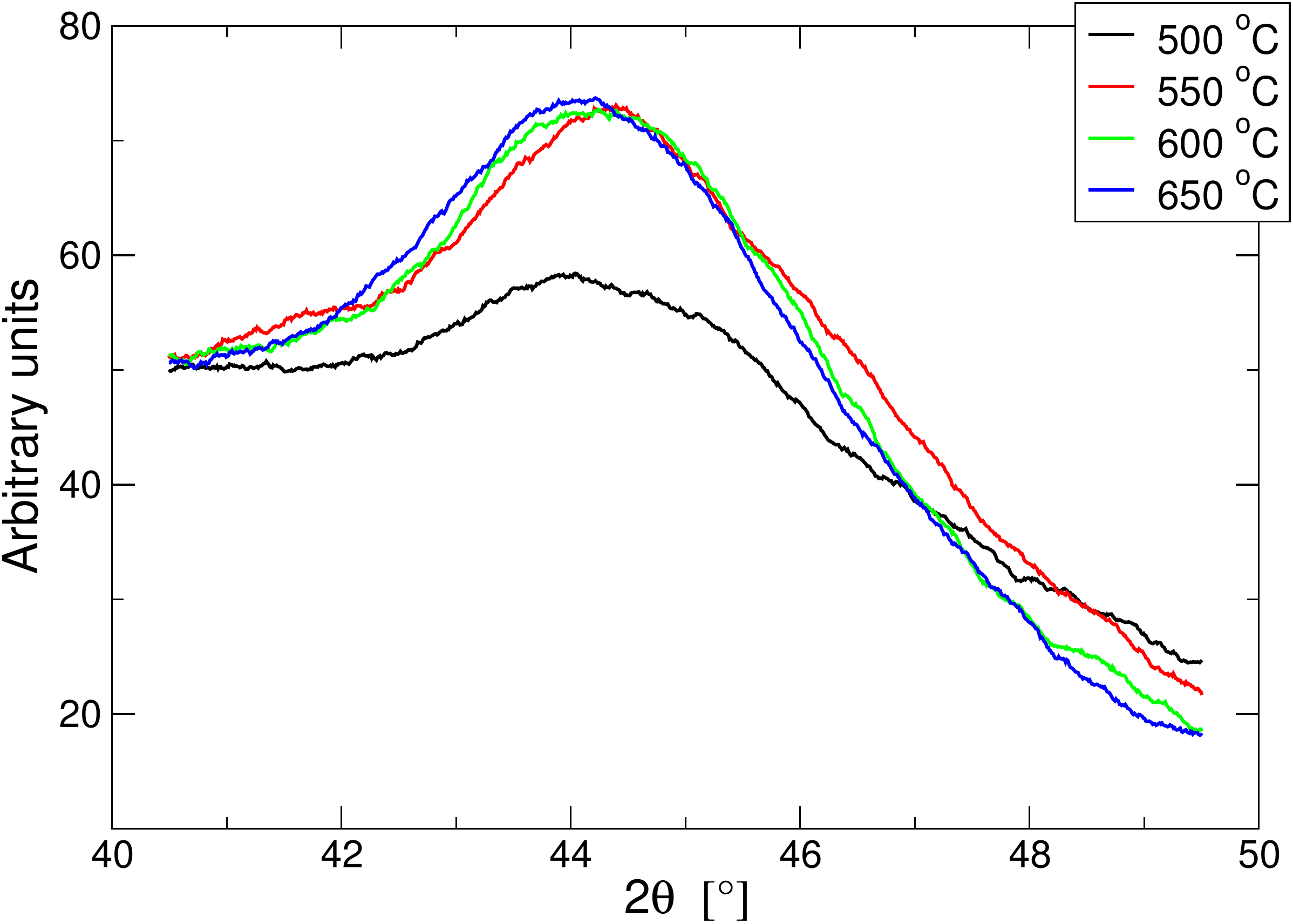}
\begin{minipage}{4.5in}\caption{\label{XRD_high_angles_fig}
Enlarged experimental XRD spectra around the peak at $\sim45^\circ$
for the g-C$_3$N$_4$ samples prepared at 500, 550, 600, and
650~$^{\circ}$C.}
\end{minipage}
\end{figure*}

\vspace{0.75in}
\begin{figure*}[h]
\includegraphics[width=0.65\textwidth]{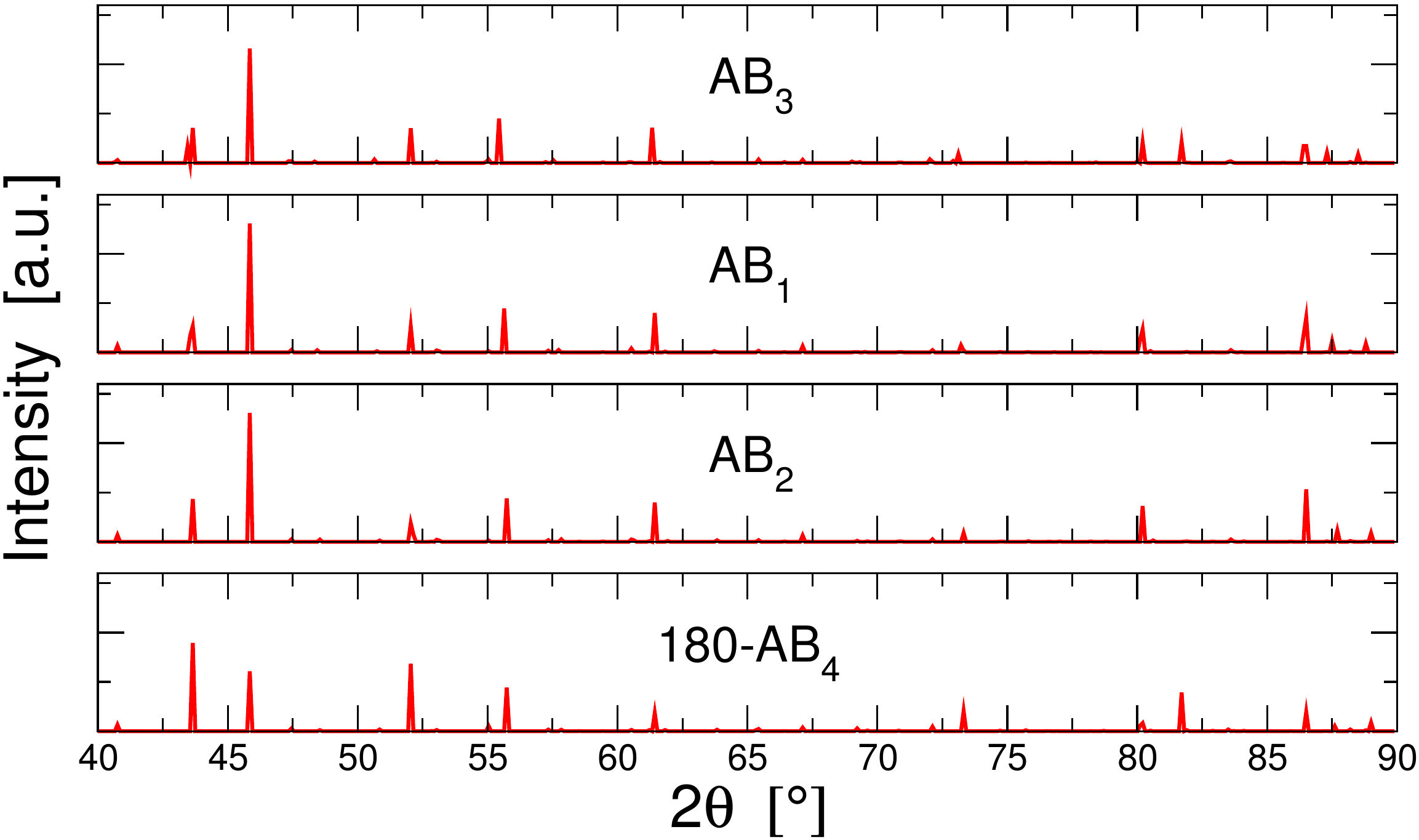} 
\begin{minipage}{4.5in}\caption{\label{XRD_theory_fig}
Calculated XRD spectra at higher angles for the four
g-C$_3$N$_4$ stacking configurations.}
\end{minipage}
\end{figure*}

\vspace*{0.25in}
\begin{figure*}[h]
\includegraphics[width=0.65\textwidth]{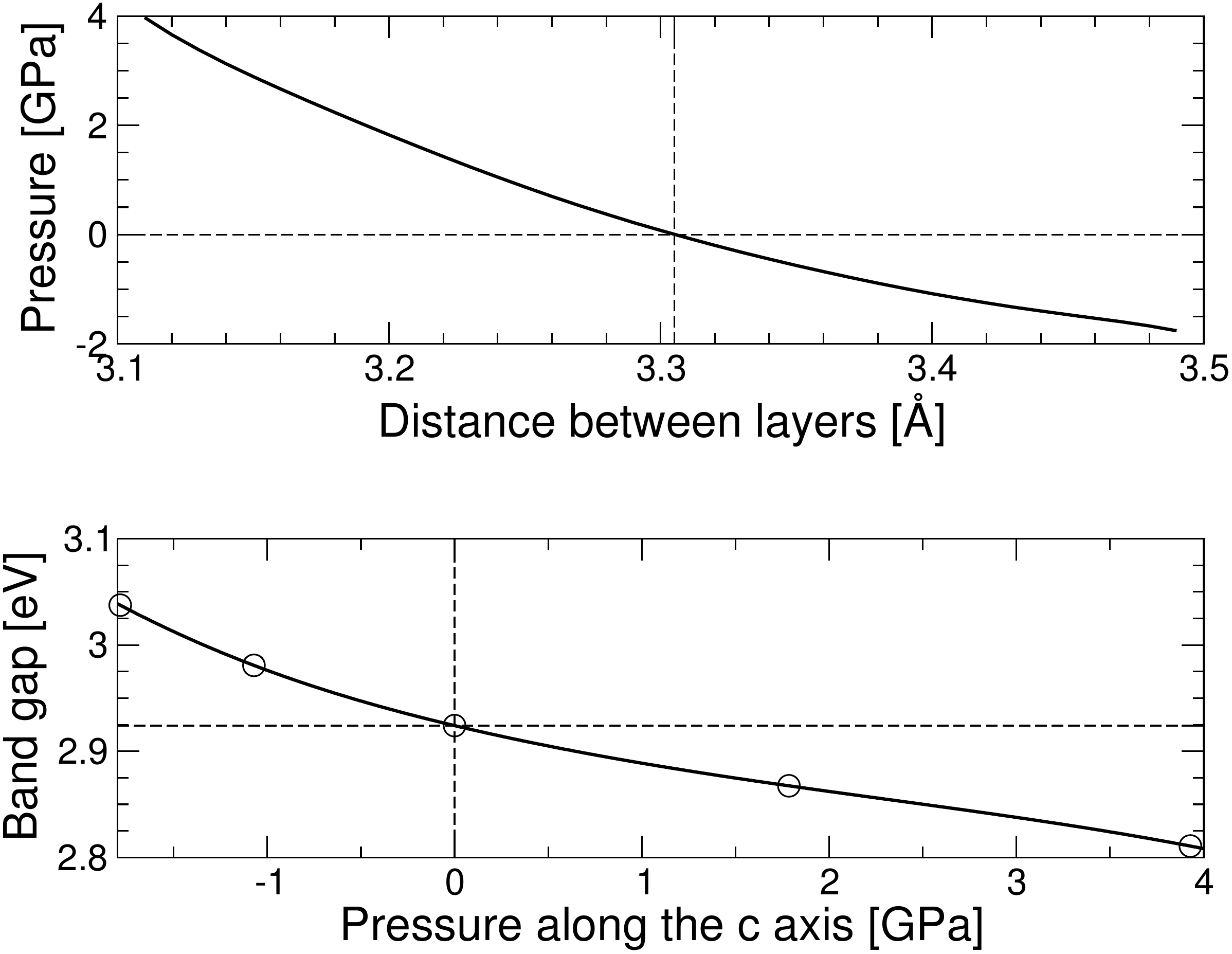}
\begin{minipage}{4.5in}
\caption{\label{Pressure_LC_BG_fig}
(top) Pressure vs.\ layer separation $d$ in the AB$_1$ stacking
of g-C$_3$N$_4$. (bottom) Band gap of AB$_1$ as a function of the
applied pressure along the $c$ axis.}
\end{minipage}
\end{figure*}

